
-------------------------

\def\oneandahalfspace{\baselineskip=16pt plus 1pt
\lineskip=2pt\lineskiplimit=1pt}

\def\np{\vfill\eject}
\def\nl{\hfil\break}

\def\nofirstpagenoten{\nopagenumbers\footline={\ifnum\pageno>1\tenrm
\hss\folio\hss\fi}}
\def\nofirstpagenotwelve{\nopagenumbers\footline={\ifnum\pageno>1\twelverm
\hss\folio\hss\fi}}
\def\leaderfill{\leaders\hbox to 1em{\hss.\hss}\hfill}


\parindent=20pt
\def\narrow{\advance\leftskip by 40pt \advance\rightskip by 40pt}

\def\nonarrower{\advance\leftskip by -40pt\advance\rightskip by -40pt}

\def\boxit#1{\vbox{\hrule\hbox{\vrule\kern3pt
        \vbox{\kern3pt#1\kern3pt}\kern3pt\vrule}\hrule}}

\def\gtorder{\mathrel{\raise.3ex\hbox{$>$}\mkern-14mu
             \lower0.6ex\hbox{$\sim$}}}
\def\ltorder{\mathrel{\raise.3ex\hbox{$<$}|mkern-14mu
             \lower0.6ex\hbox{\sim$}}}
\def\dalemb#1#2{{\vbox{\hrule height .#2pt
        \hbox{\vrule width.#2pt height#1pt \kern#1pt
                \vrule width.#2pt}
        \hrule height.#2pt}}}

\font\twelvett=cmtt12 \font\twelvebf=cmbx12
\font\twelverm=cmr12 \font\twelvei=cmmi12 \font\twelvess=cmss12
\font\twelvesy=cmsy10 scaled \magstep1 \font\twelvesl=cmsl12
\font\twelveex=cmex10 scaled \magstep1 \font\twelveit=cmti12
\font\tenss=cmss10
 
 \font\ninebf=cmbx9
\font\ninerm=cmr9 \font\ninei=cmmi9
\font\ninesy=cmsy9 
\font\eightrm=cmr8
\catcode`@=11
\newskip\ttglue
\newfam\ssfam

\def\twelvepoint{\def\rm{\fam0\twelverm}
\textfont0=\twelverm \scriptfont0=\ninerm \scriptscriptfont0=\sevenrm
\textfont1=\twelvei \scriptfont1=\ninei \scriptscriptfont1=\seveni
\textfont2=\twelvesy \scriptfont2=\ninesy \scriptscriptfont2=\sevensy
\textfont3=\twelveex \scriptfont3=\twelveex \scriptscriptfont3=\twelveex
\def\it{\fam\itfam\twelveit} \textfont\itfam=\twelveit
\def\sl{\fam\slfam\twelvesl} \textfont\slfam=\twelvesl
\def\bf{\fam\bffam\twelvebf} \textfont\bffam=\twelvebf
\scriptfont\bffam=\ninebf \scriptscriptfont\bffam=\sevenbf
\def\tt{\fam\ttfam\twelvett} \textfont\ttfam=\twelvett
\def\ss{\fam\ssfam\twelvess} \textfont\ssfam=\twelvess
\tt \ttglue=.5em plus .25em minus .15em
\normalbaselineskip=14pt
\abovedisplayskip=14pt plus 3pt minus 10pt
\belowdisplayskip=14pt plus 3pt minus 10pt
\abovedisplayshortskip=0pt plus 3pt
\belowdisplayshortskip=8pt plus 3pt minus 5pt
\parskip=3pt plus 1.5pt
\setbox\strutbox=\hbox{\vrule height10pt depth4pt width0pt}
\let\sc=\ninerm
\let\big=\twelvebig \normalbaselines\rm}
\def\twelvebig#1{{\hbox{$\left#1\vbox to10pt{}\right.\n@space$}}}

\def\tenpoint{\def\rm{\fam0\tenrm}
\textfont0=\tenrm \scriptfont0=\sevenrm \scriptscriptfont0=\fiverm
\textfont1=\teni \scriptfont1=\seveni \scriptscriptfont1=\fivei
\textfont2=\tensy \scriptfont2=\sevensy \scriptscriptfont2=\fivesy
\textfont3=\tenex \scriptfont3=\tenex \scriptscriptfont3=\tenex
\def\it{\fam\itfam\tenit} \textfont\itfam=\tenit
\def\sl{\fam\slfam\tensl} \textfont\slfam=\tensl
\def\bf{\fam\bffam\tenbf} \textfont\bffam=\tenbf
\scriptfont\bffam=\sevenbf \scriptscriptfont\bffam=\fivebf
\def\tt{\fam\ttfam\tentt} \textfont\ttfam=\tentt
\def\ss{\fam\ssfam\tenss} \textfont\ssfam=\tenss
\tt \ttglue=.5em plus .25em minus .15em
\normalbaselineskip=12pt
\abovedisplayskip=12pt plus 3pt minus 9pt
\belowdisplayskip=12pt plus 3pt minus 9pt
\abovedisplayshortskip=0pt plus 3pt
\belowdisplayshortskip=7pt plus 3pt minus 4pt
\parskip=0.0pt plus 1.0pt
\setbox\strutbox=\hbox{\vrule height8.5pt depth3.5pt width0pt}
\let\sc=\eightrm
\let\big=\tenbig \normalbaselines\rm}
\def\tenbig#1{{\hbox{$\left#1\vbox to8.5pt{}\right.\n@space$}}}
\let\rawfootnote=\footnote \def\footnote#1#2{{\rm\parskip=0pt\rawfootnote{#1}
{#2\hfill\vrule height 0pt depth 6pt width 0pt}}}
\def\tenfoot{\tenpoint\hskip-\parindent\hskip-.1cm}

\overfullrule=0pt

\def\ft#1#2{{\textstyle{{#1}\over{#2}}}}
\def\frac#1#2{{{#1}\over{#2}}}
\def\1#1{\frac1{#1}} \def\2#1{\frac2{#1}} \def\3#1{\frac3{#1}}
\def\sb#1{\lower.4ex\hbox{${}_{#1}$}}
\def\noss{\noalign{\smallskip}}
\def\noms{\noalign{\medskip}}
\def\nobs{\noalign{\bigskip}}

\def\.{\,\,,\,\,}
\def\cramp{\medmuskip = 2mu plus 1mu minus 2mu}

\cramp
\oneandahalfspace

\def\ophi{{\overline\phi}}

\rightline{CTP TAMU--59/91}
\rightline{August 1991}
\vskip 3truecm

\centerline{\bf Bosonisation of the Complex-Boson Realisations of $W_\infty$}

\vskip 2truecm
\centerline{X. Shen\footnote{$^\star$}{\tenfoot Supported in part by the U.S.
Department of Energy, under grant DE-FG05-91-ER40633} and X.J. Wang}
\bigskip\bigskip
\centerline{Center for Theoretical Physics}
\centerline{Physics Department}
\centerline{Texas A\&M University}
\centerline{College Station, TX 77843}

\vskip 2truecm

\centerline{ABSTRACT}
\bigskip

        We bosonise the complex-boson realisations of the $W_\infty$ and
$W_{1+\infty}$ algebras. We obtain nonlinear realisations of $W_\infty$ and
$W_{1+\infty}$ in terms of a pair of fermions and a real scalar. By further
bosonising the fermions, we then obtain realisations of $W_\infty$ in terms
of two scalars. Keeping the most non-linear terms in the scalars only, we
arrive at two-scalar realisations of classical $w_\infty$.

\np
\bigskip
\noindent
{\bf 1. Introduction}
\bigskip

        Since the discovery of $W_\infty$ [1] and $W_{1+\infty}$ [2], it has
become clear that there are many applications for these algebras. First
of all, as an algebra containing generators $V^i$ with spin $i+2\ge 2$,
$W_\infty$ is a leading candidate for the universal $W$-algebra [3,10,11].
Secondly, the concept of $2D$ $W$-gravity and $W$-strings as a
generalisation of ordinary $2D$ gravity and strings, which has been of much
interest recently [4,16,17], may by itself be a very fruitful idea. Thirdly
it has also been shown that $W_{1+\infty}$ algebra generates the $W_N$
constraints for the $N$-matrix model solutions for non-perturbative $2D$
quantum gravity coupled to a matter system with $c\le 1$ [5,11], and has been
further argued to be the underlying symmetry for these systems [6].

        Various free-field realisations for $W_\infty$ [7], $W_{1+\infty}$
[8,9] and super-$W_\infty$ [8] have been found. These realisations provide a
more transparent way to investigate issues concerning these algebras and
their applications. For example in Ref.\ [10], the free-fermion realisation
of $W_\infty$ at $c=-2$ has played a pivotal r\^ole in carrying out the
reduction to $W_N$ from $W_\infty$. In Ref.\ [11], the $W_{1+\infty}$
constraints were obtained from a single-scalar realisation that is the
bosonised form of the complex fermion realisation of $W_{1+\infty}$. In
Ref.\ [15,16,17], realisations of $W_\infty$ provided the matter system in
which framework the gauging of $W$ symmetries have been carried out.

       There exists a common feature among the various complex free-field
realisations of $W_\infty$, namely, the currents for these algebras are
expressed in the bilinear form of the free-fields. In the language of field
theory, they are termed linear realisations of the corresponding algebras.
On the other hand, the one real-scalar realisations of $W_\infty$ and
$W_{1+\infty}$ are different in that they are nonlinear realisations
obtained from the bosonisation of the corresponding fermion realisation.
Since one can bosonise complex bosons as well as complex fermions, it must
be possible to achieve other nonlinear realisations by bosonising the
complex-boson realisation of $W_\infty$. This is precisely what we will
accomplish in this paper.

      We shall first review various realisations of the $W_\infty$ algebras in
the literature, and illustrate their interrelationships. We then proceed to
carry out the bosonisation of the complex-boson realisations of the $W_\infty$
algebras, by the well-known procedure of Ref.\ [12]. The bosonised currents
are expressed in terms of a pair of fermions and a scalar. Bosonising the
fermions further, we obtain a realisation of $W_\infty$ in terms of two
coupled scalars. We then extract a realisation of the classical $w_\infty$
by keeping the most nonlinear terms in the scalars. Finally we conclude with
discussions.

\bigskip
\bigskip
\noindent
{\bf 2. Realisations of $W_\infty$}
\bigskip

      The first free-field realisation of the $W_\infty$ algebra was given
in Ref.\ [7]. The $W_\infty$ currents $V^i$ with spin $s=i+2\ge 2$ are
realised as bilinears of a pair of scalars $\varphi(z)$ and $\varphi^*(z)$,
with OPE
$$
\varphi^*(z)\varphi(w)\sim -\log(z-w)\ ,\eqno(1)
$$
as follows:
$$
V^i(z)=-\sum_{k=0}^{i+1}
a_k(i,\ft12):\partial^k\varphi^*(z)\partial^{i-k+2}\varphi(z):\ ,\eqno(2)
$$
where the constants $a_k(i,\alpha)$ are given by
$$
a_k(i,\alpha)={i+1\choose k}{(i+2\alpha+2-k)_k(2\alpha-i-1)_{i+1-k}\over
(i+2)_{i+1}}\ .\eqno(3)
$$
Here $(a)_n\equiv a(a+1)\cdots(a+n-1)$.

      The $W_{1+\infty}$ algebra can be realised in a similar fashion as
bilinears of a complex fermion $\psi$, with OPE
$$
{\overline \psi}(z)\psi (w)\sim{1\over{z-w}}\ .\eqno(4)
$$
There is a one-parameter family of basis in which the currents can be
realised [13,14]. The currents ${\widetilde V}^i(z)$ with spin $i+2\ge 1$ in
the
basis corresponding to an arbitrary value of the parameter $\alpha$ are
given by [14]
$$
{\widetilde V}^i(z)=\sum_{k=0}^{i+1}a_k(i,\alpha)\partial^k{\overline \psi}(z)
\partial^{i+1-k}\psi (z)\ .\eqno(5)
$$
For a generic value of $\alpha$, the central charges of the currents are not
diagonalised. When $\alpha=0$ they are diagonalised, and the corresponding
realisation of $W_{1+\infty}$ with $c=1$ was described in Ref.\ [8,9].

      Since it has been shown that $W_\infty$ can be embedded into
$W_{1+\infty}$ [13], one can obtain a realisation of $W_\infty$ from that of
$W_{1+\infty}$ by taking a suitable basis and truncating out the spin-1
current. Concretely, by taking $\alpha=\ft12$ in the above realisation (5),
one can show that ${\widetilde V}^i$ with $i\ge 0$ form a closed algebra,
namely the $W_\infty$ algebra. The central charge of this realisation is
$c=-2$, obeying the general relationship between the central charge of the
embedded $W_\infty$ algebra and that of $W_{1+\infty}$ [13].

      It is worth noting that the currents $V^i$ of $W_\infty$ (2), and
${\widetilde V}^i$ of $W_{1+\infty}$ (5) with $\alpha=0$, constitute a
realisation for the bosonic sector of the $N=2$ super-$W_\infty$ algebra,
while the fermionic currents can be constructed as mixed bilinears of $\psi$
and $\phi$ [8].

      There also exists a realisation of $W_{1+\infty}$ in terms of a
complex boson $\phi$ with OPE [14]
$$
\ophi(z)\phi(w)\sim -{1\over{z-w}}\ .\eqno(6)
$$
With respect to the stress tensor $T\equiv{\widetilde V}^0$ given by
$$
T=(\alpha+{\ft12})\partial\ophi\phi
+(\alpha-{\ft12})\ophi\partial\phi\ ,
\eqno(7)
$$
the $\phi$ and $\ophi$ have conformal spin $\ft12+\alpha$ and
$\ft12-\alpha$. In general the currents of $W_{1+\infty}$ in a basis with
parameter $\alpha$ are given by
$$
{\widetilde V}^i = \sum_{k=0}^{i+1}a_k(i,\alpha)\partial^k
\ophi\partial^{i+1-k}\phi\ .\eqno(8)
$$
In fact, this realisation has the same structure as the fermionic
realisation of $W_{1+\infty}$ with $\phi$ and $\ophi$ replacing
$\psi$ and ${\overline\psi}$ respectively. However, the central charge in
the complex-boson realisation differs from that of the complex-fermion
realisation by a sign. In particular, in the diagonalised basis at
$\alpha=0$, the central charge of this realisation is $c=-1$.

     For later convenience, we give the explicit expressions of the first
few currents in this realisation of $W_{1+\infty}$ at $\alpha=0$:
$$
\eqalign{
{\widetilde V}^{-1}&=\ophi\phi,\cr
{\widetilde V}^0&={\ft12}\partial\ophi\phi-{\ft12}\ophi\partial\phi,\cr
{\widetilde V}^1&={\ft16}\partial^2\ophi\phi
-{\ft23}\partial\ophi\partial\phi
+{\ft16}\ophi\partial^2\phi,\cr
{\widetilde V}^2&=\ft1{20} \partial^3\ophi\phi-\ft9{20}\partial^2
{\ophi}\partial\phi+\ft9{20}\partial\ophi\partial^2\phi-\ft1{20}\ophi
\partial^3\phi\ .\cr}\eqno(9)
$$

     Once again, at $\alpha={\ft12}$ one can truncate out the spin-1
current, thus obtaining a bosonic realisation of $W_\infty$ at $c=2$. In
fact, this realisation is precisely the first realisation of $W_\infty$
given in (2) [7], upon making the identifications of $\phi\rightarrow
-\partial\varphi$ and $\ophi\rightarrow\varphi^*$.

     We have reviewed two independent realisations of $W_\infty$ (and
$W_{1+\infty}$): a complex-boson realisation and a complex-fermion
realisation. They have identical structure in terms of their respective free
fields, but opposite central charges, which can be summarised as follows.
$$
\matrix{&{\hbox{Bosonic}}&{\hbox{Fermionic}}\cr\cr
        W_{1+\infty} & -1 & 1  \cr\cr
        W_\infty     & 2  & -2 \cr}
$$

     The common feature of these realisations is that the currents are
bilinears of the free fields. Therefore the transformations generated by
these currents on the free field are linear. There does exist a nonlinear
realisation of $W_\infty$ (and $W_{1+\infty}$) at $c=-2$ ($c=1$) in terms of
a real scalar $\chi$, with OPE $\chi(z)\chi(w)\sim \log(z-w)$. It can be
viewed as the bosonisation of the complex-fermion realisation given in (5)
[11], with the following identification:
$$
{\overline\psi}\equiv e^{-\chi}\ , \quad \psi\equiv e^\chi\ .\eqno(10)
$$
Recently this realisation has been argued to furnish the quantum $W_\infty$
symmetry, to which the (classical) $w_\infty$ symmetry deforms in the
process of quantisation, and has been used to extract the effective theory
of $W_\infty$ gravity [15].

    It is known in the literature that complex bosons can also be
``bosonised'' [12]. Thus it must be possible to obtain another nonlinear
realisation of $W_\infty$ by bosonising the complex-boson realisation
mentioned above. In next section we shall carry out this construction
explicitly.

\bigskip
\bigskip
\noindent
{\bf 3. Bosonisation of the Complex-boson Realisations of $W_\infty$}
\bigskip

    Let us first recall the procedure of bosonising a pair of bosons $\phi$
and $\ophi$. The novelty involved in the procedure is the necessity of
introducing a pair of fermions $\eta$ and $\xi$ in addition to a scalar
field $\sigma$ [12], which parallels the scalar $\chi$ for bosonising fermions
given in (10). These free fields satisfy the following OPEs:
$$
\eqalignno{
\sigma(z)\sigma(w)&\sim-\log(z-w)\ ,&(11)\cr
\eta(z)\xi(w)&\sim{1\over{z-w}}\ .&(12)\cr}
$$
Note that the OPE of $\sigma$ differs from that of $\chi$ by a sign. The
bosonisation is furnished by
$$
\ophi\equiv:\partial\xi e^{-\sigma}:\ ,\qquad
\phi\equiv:\eta e^\sigma:\ ,\eqno(13)
$$
where the symbol $::$ implies appropriate normal ordering.

     To proceed with the bosonisation, we expand the appropriately normal
ordered product of two bosonic operators $:\phi(w)\ophi(w):\equiv\lim_{w\to
z} \Big(\phi(w)\phi(z)-{1\over w-z}\Big)$ in powers of $(w-z)$. It reads
$$
\eqalign{
\phi(w)\ophi(z)-{1\over w-z}&=:\eta(w) e^{\sigma(w)}::\partial\xi(z)
e^{-\sigma(z)}: -{1\over w-z}\cr
\noss
&=\Big({1\over{(w-z)}^2}+:\eta(w)\partial\xi(z):\Big)\ (w-z):e^{\sigma(w)-
\sigma(z)}:-{1\over{w-z}}\cr
\noss
&=:\eta(w)\partial\xi(z):\
(w-z):e^{\sigma(w)-\sigma(z)}:+{1\over{w-z}}:e^{\sigma(w)-\sigma(z)}:-{1\over
w-z}\cr
\noss
&=\sum_{k,j=0}^\infty {{(w-z)^{k+j+1}}\over{k!j!}}\ Q^{(k)}(z)P^{(j)}(z)\cr
\noss
&\quad +\sum_{j+1}^\infty {(w-z)^{j-1}\over{j!}}\ P^{(j)}(z)\ ,\cr}\eqno(14)
$$
where $P^{(j)}$ and $Q^{(k)}$ are given by
$$
\eqalignno{
P^{(j)}(z)&\equiv :e^{-\sigma(z)}\ \partial^j\ e^{\sigma(z)}:\ ,&(15)\cr
Q^{(k)}(z)&\equiv :\partial^k\eta(z)\partial\xi(z):\ .&(16)\cr}
$$
Here we have used the fact that
$:e^{\sigma(w)}::e^{-\sigma(z)}:=(w-z):e^{\sigma(w)-\sigma(z)}:$, which
arises from eq. (11) and Baker-Campbell-Hausdorff formula. Thus this implies
that, in the limit $w\to z$,
$$
\eqalign{
:\partial_w^l\phi(w)\partial_z^m\ophi(z):\
&=\partial_z^m\partial_w^l:\phi(w)\ophi(z):\cr
\noms
&=\sum_{k,j=0}^\infty\sum_{r=0}^m\sum_{s=0}^{m-r}
{[k+j+1]_{l+r}\over{k!j!}}(-1)^r
{m\choose {r,s}} (w-z)^{k+j-l-r+1}\cr
\noms
&\qquad\quad\times\partial^sQ^{(k)}(z)\partial^{m-r-s}P^{(j)}(z)\cr
\noms
&\quad +\sum_{j=1}^\infty\sum_{r=0}^m{[j-1]_{l+r}\over{j!}}(-1)^r
{m\choose r}(w-z)^{j-l-r-1}\partial^{m-r}P^{(j)}\ ,\cr}\eqno(17)
$$
where $[a]_n\equiv a(a-1)\cdots(a-n+1)$ and ${m\choose{r,s}}$ represents the
multinomial expansion coefficient ${m!\over r!s!(m-r-s)!}$.

     Considering the coefficient of $(w-z)^0$, we obtain
$$
\eqalign{
:\partial^l\phi(z)\partial^m\ophi(z):&=\sum_{r=0}^m\sum_{s=0}^{m-r}
\sum_{k=0}^{l+r-1}(-1)^r{(l+r)!\over k!(l+r-k-1)!}{m\choose{r,s}}\cr
\noms
&\qquad\qquad\times\partial^sQ^{(k)}(z)\partial^{m-r-s}P^{(l+r-k-1)}(z)\cr
\noms
&\quad+\sum_{j=l+1}^{m+l+1}{(-)^{j-l-1}\over j}{m\choose j-l-1}
\partial^{m-j+l+1}P^{(j)}(z)\ .\cr}
\eqno(18)
$$
Using this expression and (8), we have the bosonised currents of $W_{1+\infty}$
given by
$$
{\widetilde V}^i={\widetilde V}^i_{\rm FB} +{\widetilde V}^i_{\rm B}\ ,
\eqno(19)
$$
where

$$
\eqalignno{
{\widetilde V}^i_{\rm FB} &=\sum_{k=0}^{i+1}\sum_{r=0}^k\sum_{s=0}^{k-r}
\sum_{j=0}^{i-k+r}{{(-)^r a_k(i,0) (i-k+r+1)!}\over j!(i-k+r-j)!}
{k\choose {r,s}} \partial^sQ^{(j)}\partial^{k-r-s}P^{(i+r-k-j)}\ ,&(20)\cr
{\widetilde V}^i_{\rm B}
&=\sum_{k=0}^{i+1}\sum_{j=i+2-k}^{i+2}{{(-)^{j+k-i}a_k(i,0)}\over j}
{k\choose j+k-i-2} \partial^{i+2-j}P^{(j)}\ .&(21)\cr}
$$

     Although it is guaranteed that a correct bosonisation procedure will
lead to a realisation of $W_{1+\infty}$, it is instructive to see how it
works. We shall mention some highlights in the closure of this realisation
of $W_{1+\infty}$ for the first few lower-spin currents.

     The first few currents are the following:
$$
\eqalign{
{\widetilde V}^{-1}&=\partial\sigma,\cr
\noss
{\widetilde V}^0&=-\eta\partial\xi-\ft12(\partial\sigma)^2,\cr
\noss
{\widetilde V}^1&=\partial\eta\partial\xi-\eta\partial^2\xi
+2\eta\partial\xi\partial\sigma+\ft13(\partial\sigma)^3,\cr
\noss
{\widetilde V}^2&=-\ft35\eta\partial^3\xi+\ft95\partial\eta\partial^2\xi
-\ft35\partial^2\eta\partial\xi\cr
\noss
&\quad +3\eta\partial^2\xi\partial\sigma
-3\partial\eta\partial\xi\partial\sigma-3\eta\partial\xi(\partial\sigma)^2\cr
\noss
&\quad -\ft14(\partial\sigma)^4-\ft1{10}\partial\sigma\partial^3\sigma
+\ft3{20}(\partial^2\sigma)^2\ .\cr}\eqno(22)
$$
Note that with respect to the stress tensor $T\equiv V^0$ given above,
$\eta$ and $\xi$ have conformal spin 1 and 0 respectively.

    One characteristic of this realisation is that the currents are always
bilinear in the fermions $\eta$ and $\xi$, as shown in the examples above
and the general expression given in (19)-(21). Thus it is necessary that
quartic terms of $\eta $ and $\xi$, which in principle might arise in the
OPE of two currents, be absent. This condition is indeed satisfied in the
OPEs of the first few currents, owing to the anti-commutivity of fermions
and some remarkable cancellations. For example, a term quartic in fermions
might exist in the OPE of $V^1(z)$ with $V^1(w)$, but it is zero because
both $\eta(w)\partial\xi(w)\eta(w)\partial\xi(w)=0$ and $\partial
(\eta(w)\partial\xi(w))\eta(w)\partial\xi(w)=0$. Our second example is the
OPE of $V^2(z)$ and $V^2(w)$, in which the absence of quartic fermion terms
results from both the anti-commutivity of fermions and some very nice
cancellations.

     Although the currents in this realisation consist of two distinct
parts, ${\widetilde V}^i_{\rm FB}$ and ${\widetilde V}^i_{\rm B}$ as given
in (19), and in particular ${\widetilde V}^i_{\rm B}$ involves the bosonic
field $\sigma$ only, neither one of them alone gives a realisation of
$W_{1+\infty}$. This seems to be contradictory to naive expectation, for
there exists a single-scalar realisation of $W_{1+\infty}$ [11] identical to
${\widetilde V}^i_{\rm B}$ in structure up to an overall factor $(-)^i$.
But there is a crucial difference, namely the scalar $\sigma$ has opposite
sign in the propagator (11) to that of $\chi$ , which causes ${\widetilde
V}^i_{\rm B}$ not to close. One should not be surprised, after all, since
the OPEs among ${\widetilde V}^i_{\rm FB}$ give rise to terms of the
${\widetilde V}^i_{\rm B}$ form, which supply the additional terms needed
for closure. Note, incidentally, that the central charges of ${\widetilde
V}^i_{\rm FB}$ with odd $i$ vanish.

    A different basis ($\alpha=\ft12$) can be chosen for this realisation of
$W_{1+\infty}$ in terms of $\eta$, $\xi$ and $\sigma$ for which the spin-1
current $\partial\sigma$ be truncated out. This procedure leads to a
realisation of $W_\infty$. The following are the first few currents:
$$
\eqalignno{
V^0&=-\eta\partial\xi-{\ft12}(\partial\sigma)^2-{\ft12}\partial^2\sigma,\cr
\noms
V^1&=-{\ft12}\eta\partial^2\xi+{\ft32}\partial\eta\partial\xi+2\eta\partial\xi
\partial\sigma+{\ft13}\partial^3\sigma+{\ft12}\partial\sigma\partial^2
\sigma+{\ft1{12}}\partial^3\sigma,\cr
\noms
V^2&=-{\ft15}\eta\partial^3\xi+{\ft85}\partial\eta\partial^2\xi
-{\ft65}\partial^2\eta\partial\xi\cr
\noss
&\quad +2\eta\partial^2\xi\partial\sigma-4\partial\eta\partial\xi\partial\sigma
-3\eta\partial\xi(\partial\sigma)^2-\eta\partial\xi\partial^2\sigma\cr
\noss
&\quad -{\ft14}\partial^4\sigma-{\ft12}(\partial\sigma)^2\partial^2\sigma
+\ft1{20}(\partial^2\sigma)^2-{\ft15}\partial\eta\partial^3\sigma
-\ft1{60}\partial^4\sigma,\cr
\noms
V^3&=-\ft1{14}\eta\partial^4\xi+\ft{15}{14}\partial\eta\partial^3\xi
-\ft{15}{7}\partial^2\eta\partial^2\xi+{\ft57}\partial^3\eta\partial\xi\cr
\noss
&\quad +{\ft97}\eta\partial^3\xi\partial\sigma-\ft{45}{7}\partial\eta
\partial^2\xi\partial\sigma+\ft{30}{7}\partial^2\eta\partial\xi\partial\sigma
\cr
\noss
&\quad -{\ft92}\eta\partial^2\xi(\partial\sigma)^2+\ft{15}{2}\partial\eta
\partial\xi (\partial\sigma)^2+4\eta\partial\xi (\partial\sigma)^3\cr
\noss
&\quad -\ft{27}{14}\eta\partial^2\xi\partial^2\sigma+\ft{15}{14}\partial
\eta\partial\xi\partial^2\sigma+3\eta\partial\xi\partial\sigma\partial^2\sigma
+\ft{11}{14}\eta\partial\xi\partial^3\sigma\cr
\noss
&\quad +{\ft15}(\partial\sigma)^5+{\ft12}(\partial\sigma)^3\partial^2\sigma
-\ft3{14}\partial\sigma(\partial^2\sigma)^2+\ft{11}{28}(\partial\sigma)^2
\partial^3\sigma\cr
\noss
&\quad -\ft1{28}\partial^2\sigma\partial^3\sigma+\ft1{14}
\partial\sigma\partial^4\sigma+\ft1{280}\partial^5\sigma\ .&(23)\cr}
$$
Note that the conformal spins of $\eta$ and $\xi$ remain unchanged, while
the conformal spins of $\phi$ and $\ophi$ change in correlation with those of
$e^{\sigma}$ and $e^{-\sigma}$.

\bigskip
\bigskip
\noindent
{\bf 4. $w_\infty$ from $W_\infty$ in the Complex-boson Model}
\bigskip

     It is known that algebraically, $W_\infty$ admits a contraction to
$w_\infty$ [1,2], which can be viewed as the classical limit of the quantum
$W_\infty$ algebras. From the viewpoint of field theory, this contraction
seemed to be less straightforward. In particular, it was not clear how
$w_\infty$ gravity might arise in the context of $W_\infty$ gravity
where the generating currents of $W_\infty$ were built from bilinears of
complex boson [7,14] or fermion [8,9]. This puzzle was recently resolved for
the model of complex fermion in Ref.\ [15].

     The classical $w_\infty$ gravity was first introduced in the context of
a single real scalar $\chi$ [16], whose quantisation deforms the $w_\infty$
symmetry to the quantum $W_\infty$ symmetry [15]. In the generating currents
$V^i$ of $W_\infty$, the most nonlinear terms $(\partial\varphi)^{i+2}$
furnish the original $w_\infty$ symmetry, when only single contractions are
included in the OPEs, while additional terms are quantum corrections
necessary for anomaly cancellations. To this end, a beautiful phenomenon
occurs that the generating currents of $W_\infty$ can be interpreted as the
bosonisation of the complex-fermion realisation of $W_\infty$. Therefore in
order to contract the bilinear-fermion realisation of $W_\infty$ to a
realisation of (classical) $w_\infty$, it is necessary to invoke an
intermediate step, namely the bosonisation procedure, which makes it
transparent how a classical limit of $W_\infty$ can emerge.

     Analogously one can apply this treatment to the case of the complex-boson
realisation of $W_\infty$, so as to obtain a classical limit, {\it i.e.}
$w_\infty$. The bosonisation of the complex bosons $\phi$ and $\ophi$ has
been carried out in the previous section, which sets the stage for
extracting the $w_\infty$ algebra. However, the bosonised currents given in
(19)-(21) still involve a pair of fermions $\eta$ and $\xi$ in bilinear
forms. For the same reason as mentioned above, the bilinear forms of fermion
inhibit a direct extraction of $w_\infty$. Thus we need to bosonise $\eta$
and $\xi$ also.

     Let $\rho$ be the scalar field that bosonises $\eta$ and $\xi$,
according to
$$
\eta\equiv e^\rho\ ,\qquad \xi\equiv e^{-\rho}\ ,\eqno(24)
$$
where $\rho$ satisfies OPE as follows,
$$
\rho(z)\rho(w)\sim \log(z-w)\ .\eqno(25)
$$
We find that the $Q^{(j)}$ appearing in (20) is given by
$$
Q^{(j)}={1\over{j+1}}\partial R^{(j+1)}-{1\over{j+2}} R^{(j+2)}\ , \eqno(26)
$$
where
$$R^{(j)}\equiv :e^{-\rho}\partial^j e^\rho:\ .\eqno(27)$$
Substituting this expression of $Q^{(j)}$ into (20), we have the currents
${\widetilde V}^i$ in (19) expressed in terms of two scalar fields $\rho$ and
$\sigma$. The first few currents are given by
$$
\eqalign{
{\widetilde V}^{-1}&=\partial\sigma,\cr
\noss
{\widetilde V}^0&=\ft12(\partial\rho)^2-\ft12\partial^2\rho
-\ft12(\partial\sigma)^2,\cr
\noss
{\widetilde V}^1&=-\ft23(\partial\rho)^3-\partial\sigma(\partial\rho)^2
+\partial\rho\partial^2\rho+\partial\sigma\partial^2\rho-\ft16\partial^3\rho
+\ft13(\partial\sigma)^3,\cr
\noss
{\widetilde V}^2&=\ft32(\partial\sigma)^2(\partial\rho)^2
+2\partial\sigma(\partial\rho)^3+\ft34(\partial\rho)^4\cr
\noss
&\quad -\ft32(\partial\sigma)^2\partial^2\rho
-3\partial\sigma\partial\rho\partial^2\rho
-\ft32(\partial\rho)^2\partial^2\rho\cr
\noss
&\quad-\ft3{20}(\partial^2\rho)^2+\ft12\partial\sigma\partial^3\rho
+\ft35\partial\rho\partial^3\rho-\ft1{20}\partial^4\rho\cr
\noss
&\quad -\ft14(\partial\sigma)^4-\ft1{10}\partial\sigma\partial^3\sigma
+\ft3{20}(\partial^2\sigma)^2\ .\cr}\eqno(28)
$$

      Following the strategy of Ref.\ [15], we shall keep the most nonlinear
terms in the scalars $\rho$ and $\sigma$ to obtain a realisation of the
classical limit of $W_{1+\infty}$. For example, $R^{(j)}$ (26) can be
rewritten as
$$
R^{(j)}=(\partial+\partial\rho)^j\cdot 1\ ,\eqno(29)
$$
of which the classical limit is $(\partial\rho)^j$. It follows that the
classical limit of $Q^{(j)}$ (26) is given by
$$
Q^{(j)}=-{1\over{j+2}}(\partial\rho)^{j+2}\ .\eqno(30)
$$
Similarly one can work out the classical limit of $P^{(j)}$. The classical
limit of the currents of $W_{1+\infty}$ then reads
$$\eqalign{
{\tilde v}^i&=\sum_{k=0}^{i+1}\sum_{j=0}^i{{(-)^{k+1}a_k(i,0)(i+1)!}
\over{j!(i-j)!(j+2)}}(\partial\rho)^{j+2}(\partial\sigma)^{i-j}\cr
&\quad
+\sum_{k=0}^{i+1}{{(-)^ka_k(i,0)}\over{i+2}}(\partial\sigma)^{i+2}\ .
\cr}\eqno(31)
$$
This expression can be simplified by first performing the summation over
$k$, which involves the summable $_2F_1$ hypergeometric function, and
performing the summation over $j$ for the first term in (31). The final result
reads
$$
{\tilde v}^i=(-)^i{{i+1}\over{i+2}}(\partial\rho+\partial\sigma)^{i+2}
-(-)^i\partial\sigma (\partial\rho+\partial\sigma)^{i+1}\ .\eqno(32)
$$
For concreteness, we shall present the explicit expression of the first few
currents.
$$
\eqalign{
{\tilde v}^{-1}&=\partial\sigma,\cr
{\tilde v}^0&=\ft12(\partial\rho)^2-\ft12(\partial\sigma)^2,\cr
{\tilde v}^1&=-\ft23(\partial\rho)^3-\partial\sigma(\partial\rho)^2
+\ft13(\partial\sigma)^3,\cr
{\tilde v}^2&=\ft32(\partial\sigma)^2(\partial\rho)^2
+2\partial\sigma(\partial\rho)^3+\ft34(\partial\rho)^4
-\ft14(\partial\sigma)^4\ .\cr}\eqno(33)
$$

     The generating currents given in (32) form a realisation of
$w_{1+\infty}$ in the classical sense {\it i.e.} only single contraction in
the OPE. One may explicitly check that this is indeed the case, bearing in
mind a useful lemma that $\Pi(z)\Pi(w)\sim 0$, where
$\Pi\equiv\partial\rho+\partial\sigma$. This closure in the classical sense
can also be demonstrated on the transformations of $\rho$ and $\sigma$,
which follow from the general rule for the transformations of a generic
field $\Phi$ generated by the currents ${\tilde v}^i$ with parameter $k_i$
$$
\delta\Phi=\oint {dz\over 2\pi i}k_i(z) {\tilde v}^i(z) \Phi\ .\eqno(34)
$$
For example, the first few transformations are given by
$$
\eqalign{
\delta_{-1}\rho&=0,\cr
\delta_{0}\rho&=k_0\partial\rho,\cr
\delta_{1}\rho&=-2 k_1(\partial\rho)^2-2k_1\partial\rho\partial\sigma,\cr
\delta_{2}\rho&=3k_2(\partial\rho)^3+6k_2(\partial\rho)^2\partial\sigma
+3k_2\partial\rho(\partial\sigma)^2.\cr
\nobs
\delta_{-1}\sigma&=-k_{-1},\cr
\delta_{0}\sigma&=k_0\partial\sigma,\cr
\delta_{1}\sigma&=k_1(\partial\rho)^2-k_1(\partial\sigma)^2,\cr
\delta_{2}\sigma&=-2k_2(\partial\rho)^3-3k_2(\partial\rho)^2\partial\sigma
+k_2(\partial\sigma)^3\ .\cr}\eqno(35)
$$
For completeness, the general formulae for the transformations of $\rho$ and
$\sigma$ are given by
$$\eqalign{
\delta_{k_\ell}\rho&=(-)^\ell(\ell +1)k_\ell
(\partial\rho+\partial\sigma)^\ell\partial\rho\cr
\delta_{k_\ell}\sigma&=(-)^{\ell+1}k_\ell
(\partial\rho+\partial\sigma)^\ell(\ell\partial\rho-\partial\sigma)\ .\cr}
\eqno(36)
$$
It is easy to check the following closure relations of $w_{1+\infty}$
$$
[\delta_{k_i},\delta_{k_j}]\rho=\delta_{k_{i+j}}\rho\ ,\quad
[\delta_{k_i},\delta_{k_j}]\sigma=\delta_{k_{i+j}}\sigma\ ,\eqno(37)
$$
where
$$
k_{i+j}=(j+1)\partial k_i k_j-(i+1)k_i\partial k_j\ .\eqno(38)
$$

      One can also perform the same procedure to the case of $W_\infty$ to
extract a realisation of $w_\infty$. In fact, since the rotation of basis
needed for leaving out the spin-1 current in $W_{1+\infty}$ amounts to lower
order terms only, which do not effect the most non-linear terms, we conclude
that the realisation for $w_\infty$ is identical to the one given in (32).

\bigskip
\bigskip
\noindent
{\bf 5. Conclusions and Discussions}
\bigskip

    In this paper we have bosonised the complex-boson realisation of the
$W_\infty$ algebras. We have found non-linear realisations of $W_\infty$ in
terms of a scalar field and a pair of fermions. The couplings between the
scalar and the fermions are rather subtle and crucial for the closure. We
have also shown that, by bosonising the fermions further and keeping the
most non-linear terms only, the classical limits of $W_\infty$ can be
achieved. We explicitly demonstrate the closure of these limits, namely
$w_\infty$ and $w_{1+\infty}$, in the classical sense.

    Given the fact that the classical $w_\infty$ has been realised in terms
of two scalars $\rho$ and $\sigma$, it would be interesting to find certain
group theoretic structure for them, {\it i.e.} to view $\rho$ and $\sigma$
as components of certain multiplet, as the examples in Ref.\ [16].

    It is also worth noting that the two-scalar realisation of $w_\infty$
can be used as a matter system coupled to $w_\infty$ gravity, whose
quantisation presumably deforms the symmetry structure to quantum
$W_\infty$, analogous to the quantisation of single scalar coupled to
$w_\infty$ gravity [15]. The resulting theory of two scalar fields
corresponds to the complex scalar $\varphi$ coupled to $W_\infty$ gravity
given in [17].

    It has been observed [10] that the single-scalar realisation of
$W_\infty$ at $c=-2$ can be viewed as a realisation of $W_3$ or $W_N$ under
appropriate normal ordering procedure with respect to the generating
currents. This observation has lent support to the idea of $W_\infty$ being
the universal $W$-algebra. It would be interesting to see some further
evidence from this two-scalar realisation of $W_\infty$ at $c=2$.

\vfill
\eject
\bigskip
\bigskip
\centerline{\bf ACKNOWLEDGEMENT}
\bigskip

We should like to thank C.N. Pope for careful reading of the manuscript and
helpful comments.

\bigskip
\bigskip
\centerline{\bf REFERENCES}
\bigskip
\bigskip

\item{[1]}C.N.\ Pope, L.J.\ Romans and X.\ Shen, Phys.\ Lett.\ {\bf 236B}
(1990) 173;\nl C.N.\ Pope, L.J.\ Romans and X.\ Shen, Nucl.\ Phys.\
{\bf B339} (1990) 191.

\item{[2]}C.N.\ Pope, L.J.\ Romans and X.\ Shen, Phys.\ Lett.\ {\bf 242B}
(1990) 401.

\item{[3]}A.\ Morozov,\ Nucl.\ Phys.\ {\bf B357} (1991) 619;
\nl J.M.\ Figueroa-O'Farrill, J.\ Mas and E.\ Ramos, ``Bihamiltonian
Structure of the KP Hierarchy and the $W_{KP}$ Algebra,'' Leuven preprint
KUL-TF-91/23 (May, 1991);\nl
J.M.\ Figueroa-O'Farrill and E.\ Ramos, ``Existence
and Uniqueness of the Universal $W$-algebra,'' preprint KUL-TF-91/27;
\nl F.\ Yu and Y.S.\ Wu, ``Nonlinearly Deformed $W_\infty$ Algebra and
Second Hamiltonian Structure of KP Hierarchy,'' Utah preprint (May, 1991).

\item{[4]}C.M.\ Hull, Phys.\ Lett.\ {\bf 240B} (1989) 110;\nl
K.\ Schoutens, A.\ Sevrin and P.\ van Nieuwenhuizen, Phys.\ Lett.
{\bf 243B} (1990) 245;\nl Nucl.\ Phys.\ {\bf B349} (1991) 791;\nl
E.\ Bergshoeff, C.N.\ Pope and K.S.\ Stelle, Phys.\ Lett.\ {\bf 249B} (1990)
208;\nl Y.\ Matsuo, Phys.\ Lett.\ {\bf 227B} (1989) 222;
\nl C.M.\ Hull, ``$W$-Gravity Anomalies with Ghost Loops and Background
Charges,'' preprint QMW/PH/91/15;\nl K. Schoutens, A. Sevrin and P. van
Nieuwenhuizen, ``Loop Calculations in BRST-Quantised Chiral $W_3$ gravity,''
preprint, ITP-SB-91-13;\nl C.N.\ Pope, L.J.\ Romans and K.S.\ Stelle,
``Anomaly-free $W_3$ gravity and critical $W_3$ strings,'' preprint
CERN-TH.6171/91, Phys. Lett. {\bf B} (in press), ``On $W_3$ strings,''
CERN-TH.6188/91;\nl S.R.\ Das, A.\ Dhar and
S.K.\ Rama, ``Physical properties of $W$ gravities and $W$ strings,''
preprint, TIFR/TH/91-11;\nl ``Physical states and scaling properties of $W$
gravities and $W$ strings,'' preprint,\nl TIFR/TH/91-20.

\item{[5]}E.\ Brezin and V.A.\ Kazakov, Phys.\ Lett.\ {\bf 236B} (1990)
144;\nl M.\ Douglas and S.\ Shenker, \ {\bf B335} (1990) 635;\nl D.J.\ Gross
A.\ Migdal,\ Phys. Rev. Lett. {\bf 64} (1990) 717, \ Nucl.\ Phys.\ {\bf B340}
(1990) 333; \nl M.\ Douglas, Phys. Lett. {\bf 238B} (1990) 176;
\nl R.\ Dijkgraaf, E.\ Verlinde and H.\ Verlinde,\ Nucl.\ Phys.\
{\bf B348} (1991) 435;\nl M.\ Fukuma, H.\ Kawai and R.\ Nakayama, Intl.\ J.\
Mod.\ Phys.\ {\bf A6} (1991) 1385.

\item{[6]}D.\ Gross and I.\ Klebanov, Nucl. Phys. {\bf B352} (1991) 671;\nl
N.\ Seiberg and G.\ Moore, ``From Loops to Fields in $2D$ Quantum Gravity,''
RU-91-29, YCTP-P19-91;\nl E.\ Witten, ``Ground Ring of Two Dimensional
String Theory,'' IASSNS-HEP-91/51.

\item{[7]}I.\ Bakas and E.\ Kiritsis, Nucl.\ Phys.\ {\bf B343} (1990) 185.

\item{[8]}E.\ Bergshoeff, C.N.\ Pope, L.J.\ Romans, E.\ Sezgin and X.\ Shen,
Phys.\ Lett.\ {\bf 245B} (1990) 447.

\item{[9]}D.A.\ Depireux, Phys. Lett. {\bf 252B} (1990) 586.

\item{[10]}H.\ Lu, C.N.\ Pope, X.\ Shen and X.J.\ Wang, ``The Complete
Structure of $W_N$ from $W_\infty$ at $c=-2$,'' Texas A\&M Preprint CTP
TAMU-33/91 (May, 1991).

\item{[11]}M.\ Fukuma, H.\ Kawai and R.\ Nakayama, ``Infinite-dimensional
Grassmannian structure of two-dimensional quantum gravity,'' preprint, UT-572.

\item{[12]}D.\ Friedan, E.\ Martinec and S.\ Shenker, Nucl.\ Phys.\ {\bf
B271} (1986) 93.

\item{[13]}C.N.\ Pope, L.J.\ Romans and X.\ Shen, Phys.\ Lett.\ {\bf 245B}
(1990) 72.

\item{[14]}E.\ Bergshoeff, B.\ de Wit and M.\ Vasiliev, Phys.\ Lett.\ {\bf
256B}
(1991) 199;\nl
``The structure of the super-$W_\infty(\lambda)$ algebra,'' preprint,
CERN TH-6021-91.

\item{[15]}E.\ Bergshoeff, P.\ Howe, C.N.\ Pope, E.\ Sezgin, X.\ Shen
and K.\ Stelle, ``Quantisation Deforms $w_\infty$ to $W_\infty$ Gravity,''
preprint CTP-TAMU-25/91, Imperial/TP/90-91/20, ITP-SB-91-17, Nucl. Phys.
{\bf B} (in press).

\item{[16]}E.\ Bergshoeff, C.N.\ Pope, L.J.\ Romans, E.\ Sezgin, X.\ Shen and
K.S.\ Stelle, Phys.\ Lett.\ {\bf 243B} (1990) 350.

\item{[17]}E.\ Bergshoeff, C.N.\ Pope, L.J.\ Romans, E.\ Sezgin and X.\ Shen,
Mod.\ Phys.\ Lett.\ {\bf A5} (1990) 1957.

\end